# Asymmetric motion of magnetically actuated artificial cilia

*Srinivas Hanasoge, Matthew Ballard, Peter Hesketh, Alexander Alexeev*
*George W. Woodruff School of Mechanical Engineering*
*Georgia Institute of Technology, Atlanta, GA, 30332*

**Abstract**

Most microorganisms use hair-like cilia with asymmetric beating to perform vital bio-physical processes. In this paper, we demonstrate a novel fabrication method for creating magnetic artificial cilia capable of such biologically inspired asymmetrical beating pattern essential for creating microfluidic transport in low Reynolds number. The cilia are fabricated using a lithographic process in conjunction with deposition of magnetic nickel-iron permalloy to create flexible filaments that can be manipulated by varying an external magnetic field. A rotating permanent magnet is used to actuate the cilia. We examine the kinematics of a cilium and demonstrate that the cilium motion is defined by an interplay among elastic, magnetic, and viscous forces. Specifically, the forward stroke is induced by the rotation of the magnet which bends the cilium, whereas the recovery stroke is defined by the straightening of the deformed cilium, releasing the accumulated elastic potential energy. This difference in dominating forces acting during the forward stroke and the recovery stroke leads to an asymmetrical beating pattern of the cilium. Such magnetic cilia can find applications in microfluidic pumping, mixing, and other fluid handling processes.

## 1. Introduction

Microscopic beating cilia are commonly utilized by a variety of biological organisms for such functions as nutrient transport and feeding[1,2], swimming, and sensing[3,4]. Due to their small size, hair-like cilia operate in low-Reynolds number environments, where fluid transport is dominated by viscous forces, while inertial forces are negligible. In this flow regime, cilia must perform a non-reciprocal, spatially-symmetric beating pattern in order to induce a net fluid flow[5,6]. Furthermore, biological cilia often exhibit asymmetric beating patterns that play a critical role in inducing fluid transport[7,8].

Researchers have recently developed several fabrication and actuation methods to create artificial cilia. Integration of these cilia in lab-on-chip devices has significant potential for mixing, fluid pumping, and other microscale transport processes[9–11]. Den Toonder et al.[11] demonstrated that electrostatically actuated cilia are highly promising for microfluidic mixing, although they involve a rather complicated fabrication process with multiple metal and polymer lithographic steps. Moreover, introducing electric fields to actuate artificial cilia might limit the use of this approach in applications involving biological samples.

Among different actuation approaches, magnetic cilia are particularly attractive due to the ease of actuation by a magnetic field that can be varied in time to oscillate the cilia. This method also has the advantage of not interfering with biological systems. Various sophisticated fabrication techniques to realize magnetic cilia such as roll up method[12], polymers embedded with magnetic particles[13,14], and beads self-assembly[10,15] have been developed and demonstrated. The beads self-assembly method, lacks in the control of the properties of the cilia. The number of beads in a cilium may vary and the fabrication is rather complicated involving guiding particles into trenches using acousto-optic deflectors. The mechanical properties of polymeric cilia with magnetic particles are difficult to predict as the composition of the polymer and curing temperatures affect the material properties. This method also involves complex chemistry to realize the lithographic step. In other words, an experimental realization of magnetic artificial cilia is not a trivial task. A detailed review of the different fabrication approaches to fabricate magnetically actuated cilia can be found elsewhere[16].

Researchers have developed theory and computational models to understand the dynamics of magnetic artificial cilia and demonstrate their ability to pump fluids[17]. Two and three dimensional models were used to examine the kinematics of a metallic film cilium actuated by a rotating magnetic filed[18] [19]. The net flow generated by magnetically-actuated cilia in the relation to the beating pattern and the Reynolds number was analysed[20]. A linear relation was reported between the area swept by the cilium tip and the fluid flow per cycle. Khaderi and Onck[19] have examined the effects of multiple cilia interacting in an array and discussed the effects of metachronal waves. The use of such cilia for micro-particle capture[21–24], particle transport[25], flow control[26] has been examined using computational modeling.

In this paper, we present a simple lithographic technique to realize metallic thin film cilia. Using this approach, critical parameters of the cilia such as thickness, length, and width can be controlled with high precision by changing the photomask design and the metal deposition conditions. This method follows standardized steps of thin-film microfabrication which allow for a higher degree of accuracy and reproducibility. We show that using a rotating permanent magnet, our magnetic cilia can be driven to oscillate with highly asymmetric strokes, resembling the beating of biological cilia. The asymmetric motion opens the possibility of harnessing such synthetic cilia for fluid pumping and other fluid manipulations. To the best of our knowledge experimental demonstration of the use of magnetically-actuated thin metallic films as artificial cilia is reported for the first time in this paper.

Our magnetically-oscillated artificial cilia can be utilized for various microfluidic purposes due to the ease of fabrication and actuation. It is, therefore, important to understand the kinematics of such cilia under different magnetic forcing conditions. Although, researchers have studied the mechanisms of fluid manipulation in similar systems[7,9], experimental studies characterizing the asymmetric beating cilia are still limited[13]. Here, we combine both experiments and theory to understand the kinematics of magnetically driven thin film cilia for a range of actuation condition. We use a setup with a rotating permanent magnet and visualize motion of a cilium in this plane of oscillation for a range of actuation conditions. We employ a computational model for fluid-structure interactions to validate and further examine cilium motion due to a rotating magnetic field. We establish the key parameters governing cilium motion, and identify actuation regimes enhancing asymmetry of cilium beating.

**2. Experimental and computational methodology**

*2.1 Fabrication of the nickel iron (NiFe- permalloy) cilia*

The high permeability of nickel iron permalloy (NiFe) leads to high magnetic forces, and negligible coercivity allows the cilia to re-magnetize easily in a time varying field. This in turn, ensures consistent magnetic forcing experienced by permalloy over multiple cycles of magnetic field variations, making permalloy our material of choice for fabricating artificial cilia.

We employ surface micromachining techniques in our fabrication process involving a simple two-mask lithographic process. A flow chart indicating the steps of fabrication is shown in Fig. 1a. The first step is the deposition of a sacrificial layer followed by a NiFe layer. The pattern of the cilia is imprinted using a negative photoresist (NR9 1500Py Futurex) on a glass substrate. A 40nm sacrificial layer of copper is then sputtered (Unifilm sputterer) followed by a layer of NiFe (80:20 Ni:Fe permalloy) of required thickness. Lift-off is done to remove the photoresist by dissolving it in acetone, which leaves the cilia features on the surface.

The next step is to deposit an anchor for holding the cilia on the substrate. A second lithographic step is performed to obtain these features which are deposited with 150nm of titanium. This anchor layer sticks to the glass and ensures the NiFe cilia are held on the substrate. The cilia are released by removing the sacrificial copper layer by dissolving it in 5% ammonium hydroxide which selectively etches the copper.

The stresses due to sputtering NiFe alloy curls the film away from the substrate when the supporting layer is removed. This leaves a free standing magnetic NiFe film that can be actuated by an external magnetic field. This process is simple and highly reproducible. Moreover, the fabricated devices can be stored for extended periods of time, and the sacrificial layer can be removed during the time of the experiment. The dimensions of the cilia are readily modified by changing the lithographic mask and NiFe film deposition conditions.

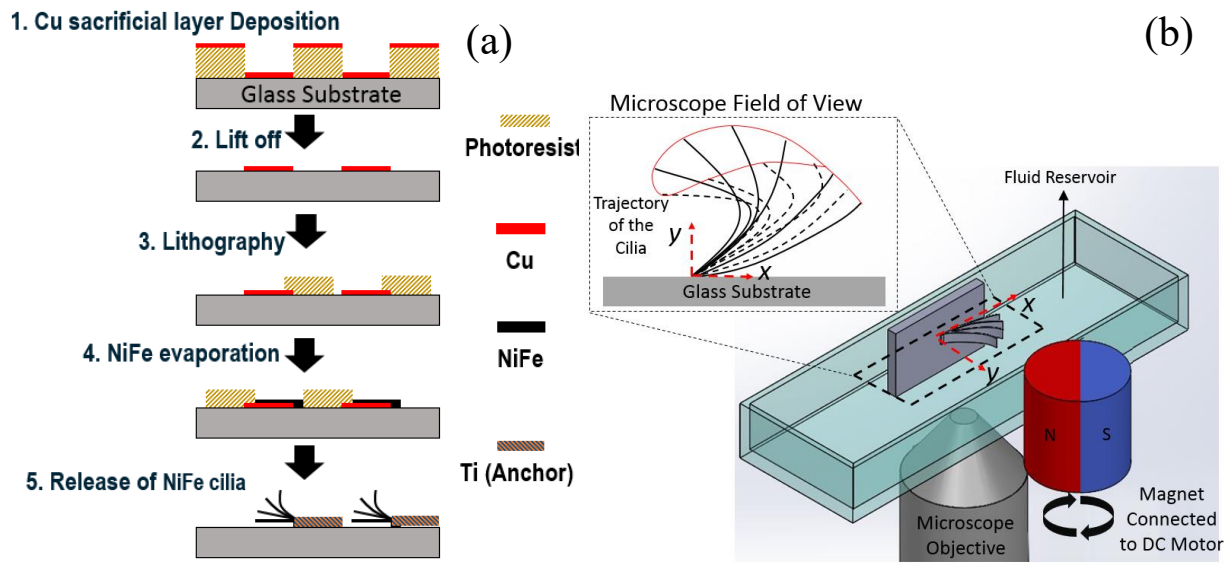

**Fig. 1** (a) Flowchart of the fabrication process showing the two steps-deposition of copper followed by NiFe layer, and titanium anchor layer. (b) Schematic of the experimental setup showing the relative positioning of the cilium on the microscope with respect to the magnet. Note the cilium oscillation plane ( $x-y$ ) is captured by the microscope.

## 2.2 Experiments

Analyzing the motion of cilia from the top view[13] does not reveal all of the important features of their motion. To better capture the cyclic motion of the cilium, we image the cilium in its plane of oscillations. To this end, we place the substrate vertically on an inverted microscope (Nikon Eclipse Ti) such that the plane of oscillations of the cilium is parallel to the microscope plane of view, $x-y$ (Fig. 1b). The cilium is actuated using a cylindrical magnet, magnetized through its diameter. The magnet rotates with its axis perpendicular to the plane of view. Since the size of the magnet (2.54 cm in diameter) is much larger than the length of the cilium (~500μm), the magnetic field experienced by the cilium is nearly uniform, i.e. the cilium experiences a uniform magnetic field that rotates in time. To capture bright-field images of the cilium motion, we use a high speed camera (Phantom Miro C110, Vision research) that is operated at 800 frames per second rate.

Trajectories of cilium oscillations are captured under varied experimental conditions. In our experiments, we vary the rotation rate of the magnet, magnetic flux density, cilium length, cilium thickness, and fluid viscosity. The magnet rotates in counter-clockwise (CCW) direction with frequencies *f* ranging from 1 to 24 Hz. The magnetic flux density is varied by changing the distance between the magnet and cilium, and measured using a gauss meter (Alpha labs, model 1140), with an average reading error of $\pm 20 Gauss$. Water ($\mu = 1 cP$), iso-propyl alcohol ($\mu = 2.1 cP$), and a 2:1 mixture of water and glycerol ($\mu = 4 cP$) are used in the experiments. Two thicknesses ($Th = 56 \pm 1 nm$, $68 \pm 1 nm$), and two lengths ($L = 480 \mu m, 430 \mu m$) of the cilium are examined. Cilium width *W* is kept constant and equal to $10 \mu m$. The modulus of NiFe is estimated to be $200 GPa$ [27]. For our experimental conditions, the Reynolds number $\text{Re} = \rho LWf / \mu$ does not exceed 0.12.

## 2.3 Numerical model

In order to simulate the dynamics of an elastic cilium in a viscous incompressible fluid, we employ a three-dimensional fully-coupled computational method for fluid-structure interactions[28–30], based on a lattice spring model (LSM)[31] for solid mechanics and lattice Boltzmann method (LBM)[32,33] for fluid dynamics. In the LSM, the cilium is modeled as a regular triangular lattice of mass points connected by harmonic stretching and bending springs, forming a continuous elastic thin plate[34]. Fluid-solid coupling between the cilium and the surrounding viscous fluid is achieved using appropriate boundary conditions[32,35]. In order to model the magnetic actuation, we impose a uniform magnetic field giving rise to a distributed moment acting on the cilium. Specifically, we locally apply magnetic moment along the cilium[36] with the magnitude that is proportional to $\sin(2\theta)$, where $\theta$ is the angle between magnetic field **B** and the local axis of the cilium[9,37,38]. The magnitude of the moment is set to match cilium deflection observed in the experiment.

## 3. Results and discussion

Microscale magnetic cilia oscillating in fluid experience forces due to the rotating magnetic field, cilium elasticity, and the viscous fluid. To better understand the effect of these forces on an oscillating cilium, we first consider an elastic soft magnetic filament subjected to a uniform magnetic field **B**. Under the action of **B**, the filament magnetizes and experiences a net moment due to this induced magnetization. This moment acts to align the poles on the filament with the direction of the field, so as to minimize its potential energy. When the filament axis is aligned with **B** (or **−B**), its magnetization is aligned with **−B** (or **B**) and the filament experiences no magnetic moment, allowing it to remain at rest in alignment with the field.

When the filament axis is misaligned with the direction of the magnetic field, the filament experiences a distributed magnetic moment, locally proportional to $\sin(2\theta)$, where $\theta$ is the angle between magnetic field **B** and the local filament axis. This local magnetic moment drives the filament to align with the direction of **B**. Due to the filament's ability to change its magnetization such that it aligns with either **B** or **−B**, the magnetic moment is maximum at $\theta$ equal to 45°, and is equal to zero when $\theta = 0°$ or $\theta = 90°$. In other words, a stationary cilium experiences a maximum magnetic force four times for every rotation of the magnetic field.

When the magnetic field rotates, it induces magnetic moments such that a free-standing filament rotates following the field rotation. If one end of the filament is anchored, the anchored portion is unable to follow the magnetic field, while the free end tends to align with the field driven by the magnetic moments. This results in bending of the elastic filament. The arising internal elastic forces act to restore the filament's initial (un-deformed) shape. Furthermore, if the filament is submerged in a viscous fluid, it also experiences a viscous force when it moves. The viscous force is proportional to the filament velocity and acts to dissipate its movement. In this limit of a low Reynolds number ($\ll 1$), the inertial forces can be neglected. Thus, the dynamic behavior of an anchored elastic magnetic filament in a viscous fluid subjected to a rotating magnetic field is determined by a balance of magnetic, elastic, and viscous forces.

We first examine the bending pattern of a cilium subjected to a magnetic field rotating with a relatively slow rotation rate (0.3 Hz) in the $x-y$ plane. Cilium motion is illustrated in Fig. 2a as a series of overlapped experimental images to show the evolution of filament positions at different time instances. These overlapped experimental images are obtained from Video 1 in the supplemental material, showing the motion of a cilium induced by CCW rotation of the magnet. In Fig. 2b, we show simulation results for cilium positions under conditions similar to the experiment presented in Fig. 2a. We find good qualitative agreement between the prediction of the computational model and experimental results. Our results are also in agreement with the simulation results reported by Khaderi and Onck[19].

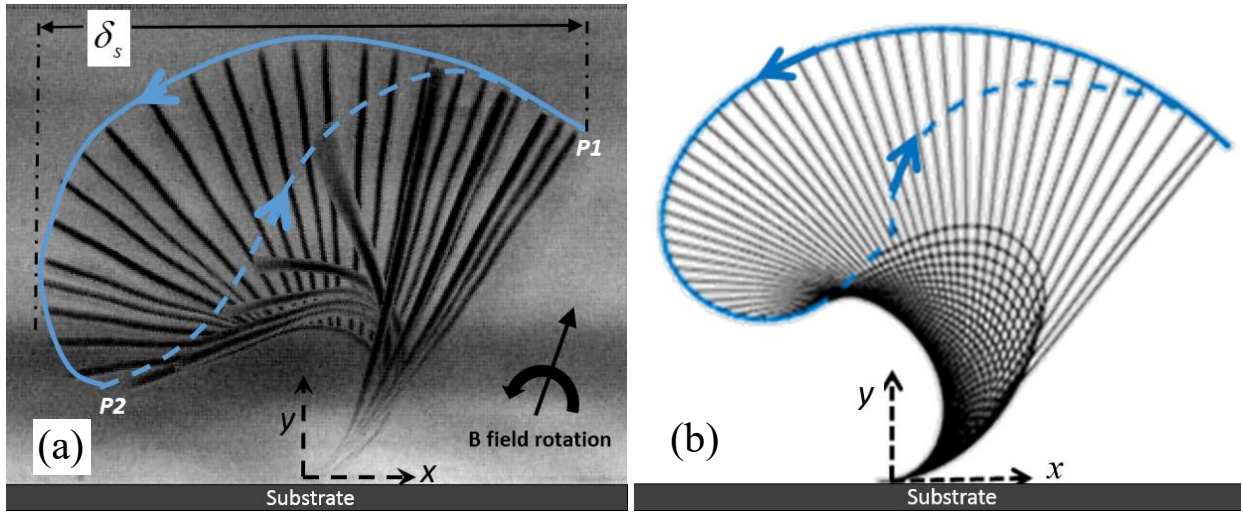

**Fig. 2** Magnetic cilium oscillation due to a CCW rotating magnetic field in the $x-y$ plane (a) Experimental snapshots of a cilium with the following parameters: $L=480\mu m$, $W=10\mu m$, $Th=55nm$, $\omega=0.3Hz$, $B=600 Gauss$, $\mu=1cP$. These parameters correspond to $Sp=3.6$ and $Mn=4.1$. Cilium positions **P1** and **P2** indicate the beginning of the forward and recovery strokes, respectively. The tip trajectories during the forward and recovery strokes are shown by the solid and dotted lines, respectively. The snapshots are obtained from Video 1, in supplemental material. (b) Trajectory of the cilium obtained using computer simulations.

The tip trajectory in Fig. 2 represents an enclosed curve with distinct paths corresponding to the cilium forward (in the direction of magnet rotation) and backward (against the direction of magnet rotation) motions. Thus, the cilium follows an asymmetric stroke where its motion differs significantly between the forward and recovery sections. The stroke asymmetry arises as a result of a difference in the forces governing the forward and backward (recovery) strokes, as we discuss below in detail.

We illustrate this magnetic forcing of elastic cilium by considering a cilium at selected positions throughout an oscillation cycle depicted in Fig. 3. The forward stroke begins when the cilium assumes the rightmost position denoted by (a) in Fig. 3. We refer the time corresponding to this position as $t = 0$. In position (a), most of the cilium is well aligned with the direction of the magnetic field, and experiences nearly zero moment, except for a short segment near the base where a bending moment arises due to the local curvature. Note that this shape is close to the equilibrium shape that the cilium assumes with no magnetic forcing, meaning that internal elastic bending moment at $t = 0$ is insignificant. As the magnetic field rotates CCW, the cilium bends such that its tip remains aligned with the field as shown for the positions (b) and (c). At the same time, the distributed magnetic moment near the base increases in magnitude and expands towards the tip. This happens due to increased cilium bending which results in greater curvature near the base. Note that at position (c), the local magnetic moment at the base changes direction and acts to enhance the bending by increasing the curvature.

As the magnetic field continues to rotate further in the counter-clockwise direction, the entire cilium experiences a significant magnetic moment as illustrated by snapshots (d) and (e) in Fig. 3. In position (e), the angle between the cilium tip and the $x$ axis reaches the maximum. We use this position to define the end of the forward stroke and the beginning of the recovery stroke. The end of the forward stroke position (e) corresponds to $t/T = 0.908$, such that the forward stroke lasts for the majority of the cycle. Here, $T$ is the total cycle time. Note that, due to the high curvature in this position, the bending magnetic moment changes sign twice along the length leading to a complex effect on cilium deformation.

Further rotation of the magnetic field causes the angle of the cilium tip to reduce. Note that the moment at the middle acts to straighten the cilium, whereas the tip remains significantly bent and experiences a moment that maintains this bending as shown for positions (e) and (f) in Fig. 3. Magnetic field rotation beyond position (f) results in a situation where the tip experiences a decreasing magnetic moment such that the elastic force exceeds it and the cilium rapidly straightens releasing the accumulated elastic energy and swiftly following positions (g) and (h). Note that in position (g) the tip is nearly normal to the direction of the magnetic field resulting in a weak bending magnetic moment. Furthermore, magnetic moment along the entire cilium, except for a small portion near the base, acts to straighten the cilium, enhancing the action of internal elastic bending moments. Thus, during this portion of the recovery stroke magnetic moments facilitates the straightening of the cilium and it moves to position (a). After the cilium returns to this position, the cycle repeats. Note that, although the instantaneous velocity of the cilia can be significant in certain parts of the recovery stroke, it is rapidly dissipated by the viscous drag keeping the inertial effects negligible[20].

An important difference between the forward and recovery strokes is related to the action of the drag force experienced by the cilium due to the surrounding viscous fluid. During the forward stroke, the cilium closely follows the magnetic field rotation and the rate of bending is proportional to the rotation rate of the field. When the rotation rate is relatively low, slow cilium velocities lead to relatively low viscous drag. In such cases, cilium motion during the forward stroke can be seen as quasi-static. That is, if the rotation of the magnetic field is stopped, the motion of the cilium will stop too and it will remain in equilibrium, in the position corresponding to a specific direction of the magnetic field. This equilibrium position is defined by a balance between the induced magnetic moment and the internal elastic force in the deformed cilium.

This quasi-static behavior can only be observed during the initial part of the recovery stroke between positions (e) and (f) in Fig. 3. After that, cilium behavior changes drastically. Between positions (f) and (a), the cilium rapidly snaps to the right due to the release of accumulated elastic energy. In this case, its velocity is defined by a balance between the internal elastic forcing and the viscous drag on the rapidly moving cilium. Indeed, the time of the forward stroke $t_f$ is more than 10 times longer that the time of the recovery stroke $t_r$. Furthermore, the time during which the cilium moves between positions (f) and (a), driven by the elastic force is only about 2% of the entire cycle, indicating a significantly larger viscous force experienced during this part of the recovery stroke.

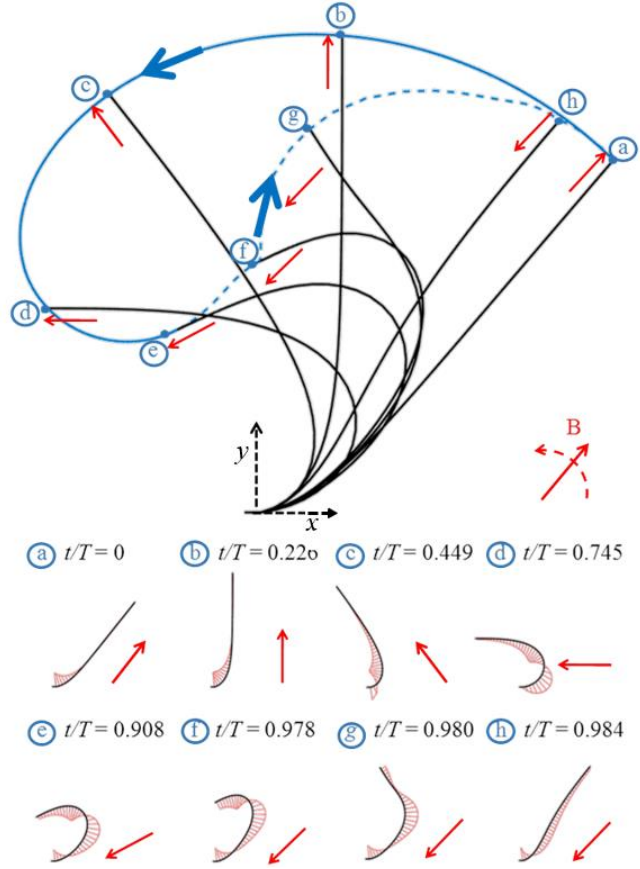

**Fig. 3** Cilium positions a-h at selected times throughout one oscillation cycle. For each cilium position, the corresponding B-field direction is given by a red arrow. The cilium tip (blue dots) traces out the tip trajectory through the forward and recovery strokes, as shown by the blue solid and dashed lines, respectively. At bottom of the figure, the magnitude and direction of the applied magnetic moment along the cilium length is shown in red as a distribution plotted normal to the local cilium axis for the corresponding positions. Note that the magnetic moment and its spatial derivative vary continuously along the length.

As discussed above, cilium motion is defined by magnetic, elastic, and viscous forces. The action of these three forces can be characterized in terms of two dimensionless numbers. We, therefore, introduce a sperm number[8,26,39] representing the ratio of viscous and elastic forces and given by $Sp = L(\omega\xi/EI)^{0.25}$. Here, $\omega$ is the oscillation frequency, $\xi = 4\pi\mu$ is the lateral drag coefficient of the cilium with $\mu$ being the dynamic viscosity of the fluid, and $EI$ is the bending rigidity. We also introduce a magnetic number[18,40] representing the ratio of magnetic and elastic forces and given by $Mn = (B^2 L^2 WTh/EI)^{0.5}$. Here, $B = |\mathbf{B}|$ is the magnitude of the magnetic flux density, $W$ and $Th$ are the cilium width and thickness, respectively, and $\mu_o$ is the permeability of free space. For our experimental setup, we estimate an average error of $\pm 0.05$ in measuring $Sp$ and $\pm 0.12$ in $Mn$. Below we examine how the motion of a cilium changes as a function of these two dimensionless numbers.

The demonstrated spatial asymmetry (Fig. 2) is critical for creating a net fluid flow[5,6]. Simulations by Khaderi et al.[18] have shown a direct dependence between the flow rate and area enclosed by the tip trajectory. It is, therefore, important to examine how the tip trajectory changes depending on the non-dimensional parameters of the system. This information is essential for the use of magnetic cilia in fluid pumping applications. In Fig. 4a, we show trajectories of the cilium tips driven at different values of $Sp$ and a constant $Mn = 4.1$. To change $Sp$ we vary the frequency of the magnetic field rotation. For low $Sp = 3.6$, the viscous drag force in the forward stroke is relatively low, allowing for large deflection of the cilium. Increasing $Sp$ increases the viscous drag during the forward stroke, reducing the magnitude of cilium deflection during the oscillation cycle. Thus, increasing $Sp$ leads to a smaller amplitude of oscillation and smaller area enclosed by the trajectory of the tip. Note that for smaller values of $Sp$ equal to 3.6 and 4.1 the tip trajectories change only slightly, indicating a limiting cycle behavior for the smaller values of $Sp$.

Fig. 4b shows the tip trajectories for different magnitudes of $Mn$ and a constant $Sp = 3.6$. Here, $Mn$ is changed by varying the magnetic flux density. We find that the cilium oscillated at a higher value of $Mn = 4.1$ exhibits a larger deflection. This is a result of a relatively large magnetic force compared to the elastic force that enables significant cilium deformation. As $Mn$ decreases, the deflection is reduced due to the lower magnetic moments, which is unable to bend the elastic cilium to achieve large deformations. Thus, reduction of $Mn$, leads to a smaller amplitude of the tip deflection and consequently smaller area enclosed by the tip trajectory.

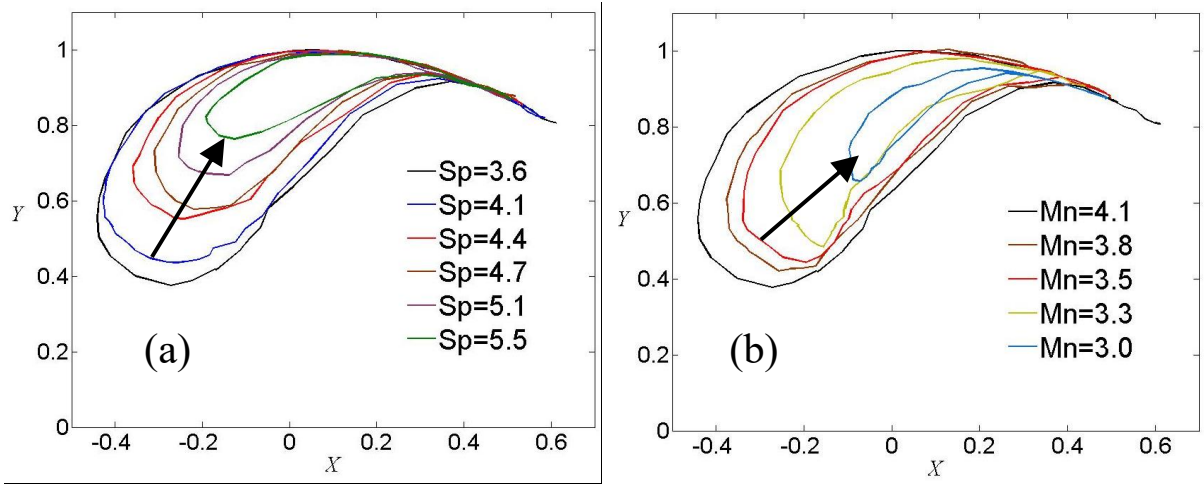

**Fig. 4** Experimental trajectories of cilium tip normalized by cilium length. (a) Tip trajectories for different values of sperm number $Sp$ varied by changing the frequencies magnetic field rotation. The arrow indicates tip trajectories with increasing $Sp$. (b) Tip trajectories for different values of magnetic number $Mn$ varied by changing **B**. The arrow shows the trajectories with decreasing $Mn$.

In the limit of low Re, fluid motion is fully defined by the kinematics of cilium stroke and its spatial asymmetry[6]. As we discussed above, the magnitude of cilium deformation during the recovery stroke is set by the balance of elastic and viscous forces. Since the viscous force is proportional to the cilium velocity, it is important to characterize the relative extent of the forward and recovery strokes that depend on the actuation regime. To this end, we define a dimensionless stroke time ratio $t_s = \dfrac{t_f - t_r}{t_f + t_r}$, where $t_f$ is the duration of the forward stroke and $t_r$ is the duraiton of the recovery stroke indicated by the cilium positions **P1** and **P2** in Fig 2a. This parameter allows us to characterize the difference in forward and recovery stroke velocities. When $t_s = 1$ the recovery velocity significantly exceeds the velocity during the forward stroke leading to a stroke with high spatial asymmetry. On the other hand, $t_s = 0$ corresponds to a stroke where the velocities of the forward and recovery strokes are equal resulting in a nearly symmetrical stroke kinematics.

In Fig 5a, we show $t_s$ as a function of $Sp$ for two values of $Mn$. In these experiments we vary $B$, $\omega$, $L$, $Th$, and $\mu$ to alter the dimensionless parameters. The values of $t_s$ collapse onto a single curve for each value of $Mn$. This confirms that the motion of an elastic magnetic cilium is indeed defined by two dimensionless numbers, $Mn$ and $Sp$. The figure also shows that $t_s$ decreases with increasing $Sp$. This can be explained by the increasing effect of viscous force damping the velocity of cilium motion, which in turn increases the time of the elasticity-driven recovery stroke. For larger $Sp$, $t_s$ converges to a constant value of about 0.25 for both $Mn$, indicating that, in this limit, the forward and recovery strokes are nearly equal. On the other hand, for smaller $Sp$ the effect of viscosity is reduced, leading to a shorter recovery time and a larger asymmetry between forward and recovery strokes, as indicated by $t_s$ approaching unity. Furthermore, a stronger magnetic force, indicated by a larger $Mn$, increases the asymmetry between the forward and recovery strokes for a given $Sp$.

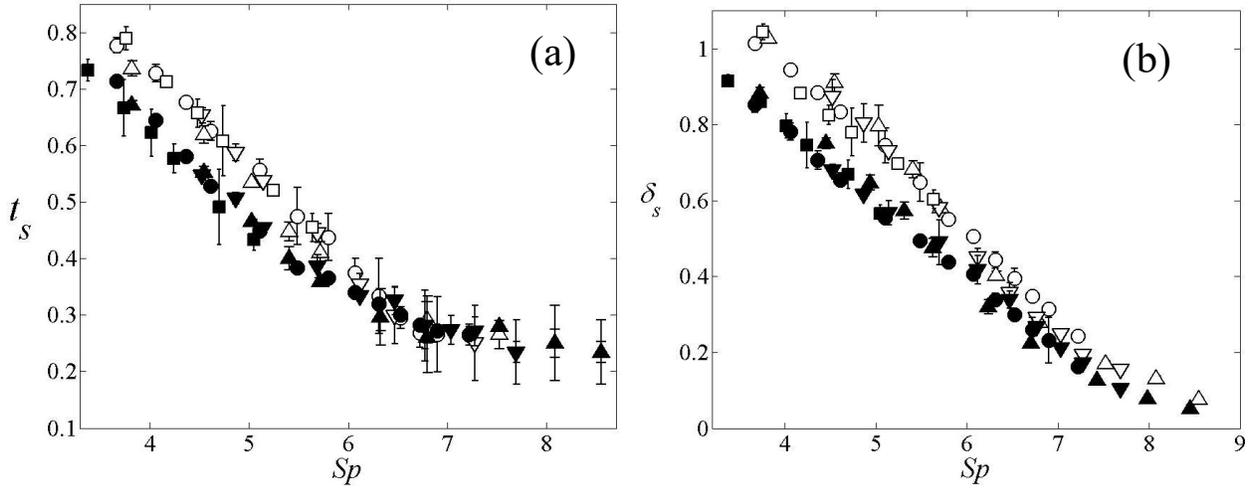

**Fig. 5** (a) Normalized time difference between forward and recovery stroke as a function of sperm number $Sp$ for different magnetic numbers $Mn$. (b) Normalized magnitude of cilium deflection as a function of sperm number $Sp$ for different magnetic numbers $Mn$. The empty markers represent $Mn = 3.6 \pm 0.12$ and the filled markers represent $Mn = 3.1 \pm 0.08$. ○: $B = 600 Gauss$, $L = 480 \mu m$, $Th = 56 nm$, $\mu = 1cP$; △: $B = 760 Gauss$, $L = 480 \mu m$, $Th = 68 nm$, $\mu = 4cP$; ▽: $B = 760 Gauss$, $L = 480 \mu m$, $Th = 68 nm$, $\mu = 2.1cP$; □: $B = 600 Gauss$, $L = 480 \mu m$, $Th = 68 nm$, $\mu = 1cP$; ●: $B = 530 Gauss$, $L = 480 \mu m$, $Th = 56 nm$, $\mu = 1cP$; ▲: $B = 650 Gauss$, $L = 480 \mu m$, $Th = 68 nm$, $\mu = 4cP$; ▼: $B = 650 Gauss$, $L = 480 \mu m$, $Th = 68 nm$, $\mu = 2.1cP$; ■: $B = 600 Gauss$, $L = 430 \mu m$, $Th = 56 nm$, $\mu = 1cP$.

To characterize the extent of the deflection of cilium, we introduce a normalized deflection amplitude $\delta_s = \delta_{max}/L$. Here, $\delta_{max}$ is the projected length of the trajectory in the $x$-direction and $L$ is the cilium length (Fig. 2a). Fig. 5b shows the variation of $\delta_s$ as a function of $Sp$ for two values of $Mn$. We find that $\delta_s$ obtained from different experiments collapse on to curves depending only on $Mn$. Further we note that $\delta_s$ decreases with increasing $Sp$. The reduction of the cilium amplitude is related to the increased effect of viscosity during the forward stroke. Indeed, viscous drag causes the cilium to lag behind rotating **B** during the forward stroke, decreasing the angle at which the cilium can no longer follow **B**. This in turn, shortens the forward stroke and decreases deflection $\delta_s$. The magnitude of cilium deflection $\delta_s$ also decreases with decreasing $Mn$. Thus, increasing $Sp$ and decreasing $Mn$ act to suppress the deflection of the cilium. We find that for larger $Sp$ the viscous force nearly completely stops cilium motion in which case $\delta_s$ approaches zero. This also agrees with a minute time difference between the forward and recovery strokes for larger values of $Sp$ shown in Fig. 5a.

We note that the Reynolds number for the data presented in Fig. 5 vary in the range between 0.006 and 0.12. The fact that the data collapse into the master curves independently of particular values of the Reynolds number supports our assumption that inertial effects play a negligible role for experimental conditions tested in our work. Note, however, that it may not be the case for larger values of Reynolds number than those used in our experiments, in which case inertial effects can influence the cilium bending pattern[20].

## 4. Conclusions

We developed a simple fabrication method for creating magnetic artificial cilia. The cilia are prepared using a two-step lithographic process by releasing a thin permalloy stripe, anchored at one end to the substrate. The method involves standardized steps for thin-film microfabrication that enables fabrication of elastic cilia with a wider range of parameters, high degree of accuracy, and reproducibility. Thus, our method can be readily adapted for the applications in lab-on-a-chip devices and can be especially useful in various biomedical assays for handling biological samples using low magnetic fields that are typically harmless. We show that when a thin permalloy cilium is actuated by a uniformly rotating magnetic field, it exhibits spatially asymmetric periodic motion. The cilium follows the rotation of the magnetic field during the forward stroke, and returns traversing a different path in the recovery stroke driven by the elastic energy accumulated in the bent cilium. We find that the time of the forward stroke which is proportional to the angular rotation rate of the magnet differs significantly from the time of the recovery stroke, set mostly by the balance between elastic and viscous forces. We characterize the relative importance of magnetic, elastic, and viscous forces in terms of dimensionless magnetic ($Mn$) and sperm numbers ($Sp$). We experimentally demonstrated the dependence of cilium kinematics on these dimensionless parameters. We show that for small values of the sperm number, the recovery stroke is significantly faster than the forward stroke, leading to asymmetric oscillations of the cilium characterized by a large area enclosed by the tip trajectory. These cilium oscillations are promising for creating spatially asymmetric motion required to induce fluid pumping and other transport processes in a low Reynolds number environment[18]. We note that our fabrication approach allows us to readily create extended arrays of multiple simultaneously beating cilia, including heterogeneous arrays of cilia with different geometries. In this work, however, we focused on examining the dynamic behavior of a single isolated cilium, thereby setting the stage for future exploration of more complex cilium systems. Multiple important effects can be investigated using magnetic cilia arrays. Computer simulations predict strong co-influence of neighboring cilia through magnetic and fluid forces[19]. Furthermore, formation of biomimetic metachronal waves has been suggested for magnetic cilia[19,41]. It has been also reported that actuation of cilia at greater Reynolds numbers can lead to dramatic changes in cilium beating pattern and ultimately flow reversal[20]. These and other effects make our magnetic elastic cilia especially attractive for a range of microfluidic applications including fluid transport and mixing, which are the subject of our future investigation.

## 5. Acknowledgments

We thank USDA NIFA grant #11317911 for financial support, and staff of Georgia Tech IEN for assistance with cleanroom fabrication.